\documentclass[twoside,reqno]{lt7-proc}
\usepackage{epsfig,cite}
\usepackage{amssymb,amsmath}
\usepackage{times}
\newcommand{\pt}{\partial_{t}}

\newcommand{\cb}{{\cal B}}

\begin{document}
\sloppy \raggedbottom
\setcounter{page}{1}

\newpage
\setcounter{figure}{0}
\setcounter{equation}{0}
\setcounter{footnote}{0}
\setcounter{table}{0}
\setcounter{section}{0}



\title{Aspects of Purely Transmitting Defects in Integrable Field Theories}

\runningheads{Corrigan}{Integrable Defects}

\begin{start}


\author{E. Corrigan}{},

\address{Department of Mathematics}{}
\address{University of York}{}
\address{York YO10 5DD, UK}{}


\begin{Abstract}
Some classical and quantum aspects of integrable defects are
reviewed with particular emphasis on the behaviour of solitons in
the sine-Gordon model.
\end{Abstract}
\end{start}


\section{Introduction}

The purpose of this talk is to review recent work exploring the
properties of `integrable defects' in 1+1 dimensional field
theories. At first sight, the idea of an integrable defect appears
preposterous because the simplest type of defect - of
$\delta$-function type, where a field is continuous across the
defect but its first spatial derivative is discontinuous - appears
to violate the property of integrability. This particular setup has
been explored numerically (see, for example \cite{Goodman2002}) and,
while there are various interesting effects associated with defects
of this type, the conclusion appears to be that one needs to look
elsewhere for an example that preserves integrability. In the
quantum domain, integrable models with defects were considered even
earlier \cite{Delf94}\cite{Konik97} and, assuming a
reasonable-looking set of relations, which should be satisfied by
the scattering matrix, reflection and transmission matrices,
integrability appeared to be incompatible (in almost all cases) with
having both reflection and transmission. There are alternative
suggestions for the relations expressing the compatibility of
reflection and transmission with the scattering matrix
\cite{Mintchev02} but they will not be pursued here for the
following reason: the types of defect considered will turn out
classically to be purely transmitting and it does not appear to be
necessary within this context to consider reflection and
transmission together.

A common `defect' in the domain of continuum mechanics is a `bore'
or  `shock' in which a field variable (the fluid velocity) has a
discontinuity (in effect modelling  rapid variations over a small
region of space), yet other quantities, such as momentum density,
are continuous. A shock is in a way more dramatic than a
$\delta$-impurity, since there is a discontinuity in the field
itself (perhaps it should be regarded as a $\delta^\prime$-defect),
yet it turns out to have a counterpart in the integrable domain.
Describing this type of defect or shock, first from a Lagrangian
view point, then exploring its classical and quantum properties, is
the purpose of this review. The main references, as far as this talk
is concerned, are the articles \cite{bczlandau}-\cite{cz07} where
many of the ideas have been introduced. There have also been
subsequent developments along similar lines and these my be found in
\cite{caudrelier07} and \cite{Gomes06}, and, as will be seen in the
quantum domain, the article \cite{Konik97} contains much of interest
for the sine-Gordon model.

\section{The setup}

The simplest situation has two scalar fields, $u(x,t)$, $x<x_0$
and $v(x,t)$, $x>x_0$, with a Lagrangian density given formally by
\begin{equation}
\label{lagrangian} {\cal L}=\theta(x_0-x){\cal L}_u +
\theta(x-x_0){\cal L}_v +\delta (x-x_0) {\cal B}(u,v)\,.
\end{equation}
The first two terms are the bulk Lagrangian densities for the
fields $u$ and $v$ respectively, while the third term provides the
sewing conditions. In principle, this term  depends on $u$, $v$ and their
various derivatives evaluated at $x_0$, but the interesting question
is how to choose ${\cal B}$ so that the resulting system remains
integrable \cite{bczlandau}.

If $u$ and $v$ are free fields, there are many ways to choose
${\cal B}$. For example,
\begin{equation}
\label{deltaimpurity} {\cal B}(u,v)=-\frac{1}{2}\,\sigma
uv+\frac{1}{2}(u_x+v_x)(u-v)\,,
\end{equation}
with standard free-field choices for the bulk Lagrangians, with
$\sigma$ a free constant parameter, leads to the following set of
field equations and sewing conditions,
\begin{equation}
\begin{array}{rcll}
(\partial^2+m^2)u&=&0\,,\quad&x<x_0\,,\\
(\partial^2+m^2)v&=&0\,,\quad&x>x_0\,,\\
u&=&v\,,\quad&x=x_0\,,\\
v_x-u_x&=&\sigma u\,,\quad& x=x_0\,,
\end{array}
\end{equation}
implying the fields are continuous with a discontinuity in the
derivative controlled by the parameter $\sigma$. This is an example of a $\delta$-impurity.
Typically, the sewing conditions at $x=x_0$ lead to reflection and
transmission and, sometimes (for $\sigma <0$) a bound state. However, if the
fields on either side have nonlinear but integrable interactions
(e.g. each is a sine-Gordon field), the $\delta$-impurity destroys
the integrability, as mentioned before \cite{Goodman2002}.

If both $u$ and $v$ are described in the bulk by integrable nonlinear wave equations,
for example they are both sine-Gordon fields, a suitable choice of
Lagrangian would be to take
\begin{equation}\label{D}
\begin{array}{rcl}
{\cal B}(u,v)&=&\frac{1}{2}(vu_t-uv_t)+{\cal D}(u,v)\,,\\[9pt]
{\cal D}(u,v)&=&
-2\biggl(\sigma\cos\frac{u+v}{2}+\frac{1}{\sigma} \cos \frac{u-v}{2}\biggr)\\
\end{array}
\end{equation}
leading to the set of equations
\begin{equation}\label{sG}
\begin{array}{rcll}
\partial^2 u&=&-\sin u\,,\quad&x<x_0\,,\\
\partial^2 v&=&-\sin v\,,\quad&x>x_0\\[6pt]
u_x&=&
v_t-\sigma \sin \frac{u+v}{2}-\frac{1}{\sigma}\sin\frac{u-v}{2}\,,\quad&x=x_0\\[9pt]
v_x&=&
u_t+\sigma\sin\frac{u+v}{2}-\frac{1}{\sigma}\sin\frac{u-v}{2}\,,\quad&x=x_0\,.
\end{array}
\end{equation}
 This setup is not at all the same as the
$\delta$-impurity because, typically,
$u(x_0,t)-v(x_0,t)\ne 0$, implying a discontinuity in the fields at $x_0$.
Clearly, the equations (\ref{sG}) describe a `defect' although the `physical'
details of the defect are hidden in the sewing conditions at $x_0$.
 Note also that the sewing conditions are strongly
reminiscent of a B\"acklund transformation, and would be a
B\"acklund transformation if they were not `frozen' at $x=x_0$
(see, for example, \cite{Backlund}). The fact the spatial derivatives
are evaluated at a specific location implies that one cannot eliminate $u$ in
favour of $v$, or vice-versa, the usual trick associated with B\"acklund
transformations. The setup is not supposed to
be obvious and was originally determined by examining
the  spin $\pm 3$ conserved charges within the sine-Gordon model
and demanding the energy-like combination was preserved \cite{bczlandau}.
That the setup is integrable
is indicated strongly in \cite{bczlandau} and \cite{bcztoda} by constructing
Lax pairs using techniques
similar to those described in \cite{bcdr} applicable to field theories
restricted to a half-line.

Since the sewing conditions (\ref{sG}) are local, it is clear there
could be many defects, with parameters $\sigma_i$,  located at $x_i$
along the $x$-axis.

As an exercise, it is worth looking at the linear approximation to
the sine-Gordon equations and defect sewing conditions \eqref{sG},
and investigating what happens to a plane wave as it approaches the
defect. Perhaps surprisingly, there turns out to be no reflected
component and the transmitted wave collects an additional phase;
there is no bound state for any value of $\sigma$.

\subsection{Energy and momentum}

Although the first analysis of this problem concentrated on
preserving the energy-like combination of spin $\pm 3$ charges, it
was natural to go back and consider energy and momentum, which are
combinations of spin $\pm 1$ charges. Despite
the conditions \eqref{sG} generally implying a discontinuity in the fields at
the defect it might still be the case that the defect could exchange
both energy and momentum with the defect itself. If so, the defect would
indeed be an integrable analogue of a shock.

Clearly, time translation invariance is not violated by the defect and
therefore there is a conserved energy, which will include a
contribution from the defect itself. On the other hand, space
translation is violated by a defect located at a specific point
and therefore momentum would
not be expected to be conserved, even allowing for a contribution
from the defect. This aspect  was also treated in \cite{bczlandau} and
investigated there in a general context
using the quantity ${\cal D}(u,v)$ appearing in (\ref{D}). Surprisingly,
including a suitably chosen defect contribution does lead to a conserved
momentum (though this would not be possible for a $\delta$-impurity).

The momentum carried by the fields on
either side of the defect is given by
\begin{equation}\label{momentum}
P=\int_{-\infty}^{x_0} dx\,u_xu_t+\int_{x_0}^\infty dx\,v_xv_t
\end{equation}
and it is necessary to check the extent to which this fails to
be conserved. Using the defect conditions coming from (\ref{D}),
\begin{equation}\label{}
u_x=v_t-\frac{\partial{\cal D}}{\partial u}\,,\quad
v_x=u_t+\frac{\partial{\cal D}}{\partial v}\,,\quad\mathrm{at}\;\,
x=x_0\,,
\end{equation}
one finds
\begin{equation}\label{Pdot}
\dot P=\left[-v_t \frac{\partial{\cal D}}{\partial u} - u_t
\frac{\partial{\cal D}}{\partial v}-V(u)+V(v)
+\frac{1}{2}\left(\frac{\partial{\cal D}}{\partial u}\right)^2 -
\frac{1}{2}\left(\frac{\partial{\cal D}}{\partial v}\right)^2
\right]_{x_0}.
\end{equation}
In this expression, the fields on either side of the defect have
been allowed to have more general potentials (and a further
generalization would have a different potential for each field).
Clearly, (\ref{Pdot}) is not generally a total time-derivative of a
functional of the two fields evaluated at $x_0$ although that is
what it is required to be in order to be able to construct a
momentum contribution located at the defect. However, it will be
provided the first two terms are a total time derivative and the
other terms exactly cancel. In other words, the quantity $\cal{D}$
and the bulk potentials must satisfy:
\begin{equation}\label{Dequations}
\frac{\partial^2{\cal D}}{\partial u^2}=\frac{\partial^2{\cal
D}}{\partial v^2}\,, \qquad \frac{1}{2}\left(\frac{\partial{\cal
D}}{\partial u}\right)^2 - \frac{1}{2}\left(\frac{\partial{\cal
D}}{\partial v}\right)^2 = V(u)-V(v)\,.
\end{equation}
This set of conditions is satisfied by the sine-Gordon defect
function (\ref{sG}). However, there are other solutions too, including
cases with several scalar fields \cite{bczlandau,bcztoda}. An
intriguing feature is that the requirements of integrability, as expressed
by analysing higher spin charges or by constructing suitable Lax
pairs, are the same as the requirements necessary for a modified conserved
momentum, at least for the sine-Gordon model. There is some evidence that not
all the higher spin conserved charges are preservable for more general
affine Toda field theories (see \cite{bcztoda} where an example of an even
spin charge is analysed in detail).

\subsection{Classical scattering of solitons}

As mentioned previously, it is easy to verify that the free-field limit
of the sine-Gordon defect setup, provided by
\begin{equation}\label{freelimit}
{\cal D}(u,v)\rightarrow \frac{\sigma}{4}(u+v)^2+\frac{1}{4\sigma}(u-v)^2,
\end{equation}
together with quadratic limits of the bulk potentials, leads to
conditions describing a purely transmitting defect. And, in any case
it is straightforward to check directly that \eqref{freelimit}
satisfies the above conditions \eqref{Dequations}. Given this, it is
natural to ask what happens to solitons in the full nonlinear
sine-Gordon model as they encounter a defect. (For a treatise on
solitons, see for example \cite{Scott73}.

A soliton travelling in the positive $x$ direction (rapidity
$\theta$) in the absence of any defect is given by the expression
\begin{equation}\label{soliton}
    e^{i u/2}=\frac{1+i E}{1-i E}
\end{equation}
where
\begin{equation}
E=e^{ax+bt+c}\,,\quad a=\cosh\theta\,,\quad
b=-\sinh\theta\,,\quad \hbox{with $e^c$ real}.
\end{equation}
The expression \eqref{soliton} provides a real solution to the
sine-Gordon equation that smoothly interpolates between
$u(t,-\infty)=0$ and $u(t, \infty)=2\pi$. If on the right hand side
of \eqref{soliton} $E$ is replaced by $-E$ the resulting solution
smoothly interpolates between $u(t,-\infty)=2\pi$ and $u(t,
\infty)=0$, and is called an anti-soliton. (Note that shifting $u$
everywhere by an integer multiple of $2\pi$ is an invariance of the
sine-Gordon equation and the soliton and anti-soliton could be
regarded alternatively as interpolating adjacent `constant
solutions', or `vacua'.)

If there is a defect then it turns out that a soliton meeting the defect will generally
pass right through (that is, it is purely transmitted) and the solutions on either
side of the defect will be given by,
\begin{equation}
e^{i u/2}=\frac{1+i E}{1-i E}\,,\quad  x<x_0\,;\qquad e^{i
v/2}=\frac{1+i zE}{1-i zE}\,,\quad  x>x_0,
\end{equation}
where $z$ is determined by the sewing conditions. Trying to allow
more general solutions (for example, to include a reflected and,
therefore, additional soliton at late times) does not satisfy the
defect sewing conditions. The defect conditions (\ref{sG}) are
satisfied requires the parameter $z$ to be
\begin{equation}\label{z}
z=\frac{e^{-\theta}+\sigma}{e^{-\theta}-\sigma}\equiv \coth
\left(\frac{\eta-\theta}{2}\right),
\end{equation}
where $\sigma=e^{-\eta}$.

One of the remarkable properties of sine-Gordon solitons is that in
the absence of any defect two solitons originally ordered with the
slower ahead of the faster will, over time, reorder to a situation
in which the faster is eventually ahead. This property has been
known since the 1960s (see, for example, \cite{Scott73}). This
scattering process is registered by a positive `delay', given by
 $z^2$ if a soliton of rapidity $\theta$ passes another of
rapidity $\eta$.

However, when a soliton encounters a defect the quantity $z$ given
by \eqref{z} may change sign, implying  a soliton might, depending
on the sign of $\theta-\eta$, convert to an anti-soliton, be
delayed, or even absorbed. In the latter case, the defect would gain
a unit of topological charge in addition to storing the energy and
momentum of the soliton; in the former, the defect gains (or loses)
two units of topological charge, as measured by the difference
$v(t,x_0)-u(t,x_0)$. In other words, a discontinuity at the location
of the defect would be created in these cases. Because the defect
potential has period $4\pi$, all evenly charged defects have
identical energy--momentum, as do all oddly charged defects.

A fascinating possibility associated with this type of defect (if it
could be realized in practice) would be the capacity to control
solitons. For example, in ref\cite{cz04} it was pointed out how
solitons together with a defect might be used to mimic a Toffoli
universal logic gate. This idea is theoretically viable using two
basic ingredients in addition to the above features. First, it
should be agreed that a soliton represents the bit `1' and an
anti-soliton represents `0'; second, there needs to be a feedback
mechanism that increases the defect parameter if a soliton passes,
but not if an anti-soliton passes. The Toffoli gate manipulates
three incoming bits in such a way that the third bit is flipped
providing the first two bits were set to `1' but not otherwise. The
Toffoli gate is universal because it can be used in combinations to
create all other gates. Of course, this cannot be other than a cute
notion unless a physical situation is found where the conditions of
the integrable defect are met.

Several defects affect progressing solitons independently  and
several solitons approaching a defect (inevitably possessing
different rapidities) are affected independently, with at most one
of the components being absorbed (and this fact was already alluded
to in the previous paragraph). Notice, too, that the situation is
not time-reversal invariant owing to the presence of explicit time
derivatives in eqs(\ref{sG}). Also, one might imagine that starting
with an odd charged defect energy--momentum conservation would
permit a single soliton to emerge. However, within the classical
picture this cannot happen because there is nothing to determine the
origin of time for the emerging soliton (that is, nothing determines
the time at which the decay would occur). Associated with this is an
interesting question for quantum mechanics. In the quantised theory
one might expect to calculate, starting with a suitably prepared
defect carrying energy and momentum, a probability for its decay at
any specified time. In a sense, this is what happens but to see how
one needs to explore the properties of the transmission matrix
within the sine-Gordon quantum field theory (see the next section).

As a final comment, for which there is no space for details, it is
worth pointing out that it is possible to generalise the setup to
allow for moving defects \cite{bczsg05}. In the classical picture a
defect located at $x=p(t)$ satisfies $\ddot{p}=0$. In other words it
may move with constant speed or remain at rest; effectively, the
defect experiences no forces as a consequence of its interaction
with the fields on either side of it. Two defects moving with
constant but differing speeds will change places eventually, or
`scatter', if the slower defect is ahead to start with. However, a
soliton encountering the pair will be influenced in a manner
independent of the ordering of the defects. In a way, it is natural
that the defect experiences no force: if it did there would need to
be an explanation of its mass. On the other hand, in  the quantum
domain the mass could be generated by quantum effects and the
scattering of two defects might turn out to be interesting. So far,
although there is a candidate for the S-matrix to describe the
scattering of two defects \cite{bczsg05} it is not yet  clear that
it can satisfy all the additional requirements of unitarity,
crossing, and so on.

\section{Quantum picture}

The sine-Gordon quantum field theory has been well-studied for many
years and much is known about the bulk theory and the theory
confined to a half-line; rather less is known about the theory
confined to an interval. As far as this talk is concerned the
essential ingredient needed from the bulk theory is Zamolodchikov's
soliton-soliton S-matrix, which will be used in the following form
(for example, see the review \cite{Zams79}),
\begin{equation}\label{Smatrix}
    S_{kl}^{mn}(\Theta)=\rho(\Theta)\left(%
\begin{array}{cccc}
  a(\Theta)& 0 & 0 & 0\\
  0 & c(\Theta)& b(\Theta) & 0 \\
  0 & b(\Theta) & c(\Theta) & 0\\
  0 & 0 & 0 & a(\Theta)\\
\end{array}%
\right),
\end{equation}
where $k,l$ label the incoming particles and $m,n$ label the
outgoing particles in a two-body scattering process, with the
particles labelled $k,n$ having rapidity $\theta_1$, and the
particles labelled $l,m$ having rapidity $\theta_2$. The various
pieces of the matrix are defined by
\begin{equation}
\quad a(\Theta)=\frac{qx_2}{x_1}-\frac{x_1}{qx_2},\quad b(\Theta)=
\frac{x_1}{x_2}-\frac{x_2}{x_1},\quad c(\Theta)=q-\frac{1}{q},
\end{equation}
with
\begin{equation}\label{}
    \Theta=\theta_1-\theta_2,\quad q=e^{i\pi\gamma},\quad x_p=e^{\gamma\theta_p}.
\end{equation}
In this notation the crossing property of the S-matrix is
represented by
\begin{equation}\label{Scrossing}
    S_{k\,l}^{m\,n}(i\pi-\Theta)=S_{k\,\, -m}^{-l\, \,n}(\Theta),
\end{equation}
with the diagonal elements $S_{+-}^{+-}(\Theta)$ and
$S_{-+}^{-+}(\Theta)$ crossing into themselves. The overall factor
$\rho(\Theta)$ will be needed later and is given by:
\begin{equation}\label{Smatrixrho}
    \rho(\Theta)=\frac{\Gamma(1+i\gamma\Theta/\pi)\Gamma(1-\gamma -i\gamma\Theta/\pi)}{2\pi i}
    \prod_{k=1}^\infty\, R_k(\Theta)R_k(i\pi -\Theta),
\end{equation}
where
\begin{equation}\label{}
    R_k(\Theta)=\frac{\Gamma(2k\gamma +i\gamma\Theta/\pi)\Gamma(1+2k\gamma +i\gamma\Theta/\pi)}
    {\Gamma((2k+1)\gamma +i\gamma\Theta/\pi)\Gamma(1+(2k-1)\gamma +i\gamma\Theta/\pi)}.
\end{equation}

Note, the conventions adopted by Konik and LeClair \cite{Konik97}
have been used. Therefore, in particular, the coupling $\gamma$ in
terms of the Lagrangian coupling $\beta$ is defined by
\begin{equation}\label{gamma}
    \frac{1}{\gamma}=\frac{\beta^2}{8\pi-\beta^2}.
\end{equation}
Where $\hbar=1$ and the conventions are those associated with the
bulk Lagrangian
\begin{equation}\label{sGbulk}
    {\cal L}=\frac{1}{2}\left((\partial_t u)^2-(\partial_x u)^2\right) - \frac{m^2}{\beta^2}
    (1-\cos\beta u).
\end{equation}

\subsection{The transmission matrix}

Following the remarks in the previous section, concerning the
classical scattering of a soliton by a defect, where the topological
charge of the defect will typically change by two units at a time,
one expects two types of transmission matrix, one of them, $^{\rm
even}T$, referring to even-labelled defects  and the other, $^{\rm
odd}T$, referring to odd-labelled defects. On the assumption that
the lowest energy state corresponds to no discontinuity at all, the
former is expected to be unitary while the latter is expected to be
related to the absorption of a soliton, and not expected to be
unitary since the excited defect is expected to decay quantum
mechanically. Rather,  $^{\rm odd}T$ is expected to be related (via
a bootstrap principle) to a complex bound state pole in $^{\rm
even}T$. In fact this is precisely what happens. It is worth
remarking that  the relevant transmission matrices were in fact
described by Konik and LeClair some time ago \cite{Konik97},
although no distinction between odd and even labels was made, nor
did those authors note the complex bound state.

It is convenient to use roman labels to denote soliton states
(taking the value $\pm 1$), and greek labels to label the charge on
a defect. Then, assuming topological charge is conserved in every
process, it is expected that both transmission matrices will satisfy
`triangle' compatibility relations with the bulk $S$-matrix, for
example:
\begin{equation}
\label{STT} S_{ab}^{cd}(\theta_1-\theta_2)\,
T_{d\alpha}^{f\beta}(\theta_1)\,T_{c\beta}^{e\gamma}(\theta_2)=
T_{b\alpha}^{d\beta}(\theta_2)\,T_{a\beta}^{c\gamma}(\theta_1)\,
S_{cd}^{ef}(\theta_1-\theta_2)\,.
\end{equation}
Here, it is supposed the solitons are travelling along the positive
$x$-axis ($\theta_1>\theta_2>0$). The bulk $S$-matrix depends on the
bulk coupling $\beta$ via the quantity $\gamma=8\pi/\beta^2 -1$, and
the conventions used are those adopted in \cite{bczsg05}. The
equations (\ref{STT}) are well known in many contexts involving the
notion of integrability (see \cite{Jimbo89}), but were discussed
first with reference to  defects by Delfino, Mussardo and Simonetti
\cite{Delf94}; if the possibility of reflection was to be allowed an
alternative framework (such as the one developed by Mintchev,
Ragoucy and Sorba \cite{Mintchev02}), might be more appropriate.
However, in the present case the defect is expected to be purely
transmitting.

The solution (for general $\beta$, and for even or odd labelled
defects --- note the labelling is never mixed by (\ref{STT})) --- is
given by
\begin{equation}
\label{KL}
T_{a\alpha}^{b\beta}(\theta)=f(q,x)\left(\begin{array}{cc}
\nu^{-1/2}Q^\alpha\delta_\alpha^\beta &
q^{-1/2}e^{\gamma(\theta-\eta)}\delta_\alpha^{\beta-2}\\[5pt]
q^{-1/2}e^{\gamma(\theta-\eta)}\delta_\alpha^{\beta+2}&
\nu^{1/2}Q^{-\alpha}\delta_\alpha^\beta\\
\end{array}\right).
\end{equation}
The solution was derived in \cite{bczsg05} and found to be
essentially unique and equivalent to the earlier result of Konik and
LeClair. A block form has been adopted with the labels $a$, $b$
labelling the four block elements on the right hand side, and where
$\nu$ is a free parameter, as is $\eta$ (to be identified with the
defect parameter introduced in the previous section), and
\begin{equation}
q=e^{i\pi\gamma}\,,\quad x=e^{\gamma\theta}\,,\quad
Q^2=-q=e^{4\pi^2i/\beta^2}\,.
\end{equation}
In addition, $^{\rm even}T$ is a unitary matrix (for real $\theta$),
and both types of transmission matrix must be compatible with
soliton--anti-soliton annihilation as a virtual process. Here, the
thinking is equivalent to that of Konik and LeClair, but expressed
rather differently. This transmission matrix represents a process
with the incoming particle meeting the defect from the left and the
process with a particle arriving from the right will be different
(though related by crossing). These two requirements place the
following restrictions on the overall factor for the even
transmission matrix, $^{e}f(q,x)$, (and henceforth $e\equiv $\lq
${\rm even}$\rq):
\begin{equation}
\left\{\begin{array}{l}
{}^{e}\bar f(q,x)={^{e}f}(q,qx)\,,\\[5pt]
{}^{e}f(q,x)\;{^{e}\!f}(q,qx)\left(1+e^{2\gamma(\theta-\eta)}\right)=1\,.
\end{array}\right.
\end{equation}
These do not determine $^{e}f(q,x)$ uniquely but the `minimal'
solution determined by Konik--LeClair has
\begin{equation}
\label{KLf} {}^{e}f(q,x)=
\frac{e^{i\pi(1+\gamma)/4}}{1+ie^{\gamma(\theta-\eta)}}\,
\frac{r(x)}{\bar r(x)}\,,
\end{equation}
with ($z=i\gamma(\theta-\eta)/2\pi$),
\begin{equation}
r(x)=\prod_{k=0}^\infty\,
\frac{\Gamma\left(k\gamma+\frac{1}{4}-z\right)\,\Gamma\left((k+1)\gamma+\frac{3}{4}-z\right)}
{\Gamma\left((k+\frac{1}{2}\right)\gamma+\frac{1}{4}-z)\,
\Gamma\left((k+\frac{1}{2})\gamma+\frac{3}{4}-z\right)}\,.
\end{equation}
It is worth noting that the apparent pole in (\ref{KLf}) at
$1+ie^{\gamma(\theta-\eta)}=0$ is actually
cancelled by a pole at the same location in $\bar r(x)$. However, there is another pole
at
\begin{equation}
\theta=\eta -\frac{i\pi}{2\gamma} \rightarrow \eta\;\;{\rm
as}\;\;\beta\rightarrow 0\,,
\end{equation}
uncancelled by a zero, and this does actually represent the
expected unstable bound state alluded to in the first section.

Several brief remarks are in  order. It is clear, on examining
(\ref{KL}), that the processes in which a classical soliton would
inevitably convert to an anti-soliton are clearly dominant even in
the quantum theory, yet suppressed if a classical soliton is merely
delayed. This much is guaranteed by the factor
$e^{\gamma(\theta-\eta)}$ appearing in the off-diagonal terms. A
curious feature is the different way solitons and anti-solitons are
treated by the diagonal terms in (\ref{KL}). They are treated
identically by the bulk $S$-matrix yet one should not be surprised
by differences showing up in the transmission matrix since the
classical defect conditions (\ref{sG}) do not respect all the usual
discrete symmetries. Indeed, the dependence of the diagonal entries
on the bulk coupling can be demonstrated to follow from  the
classical picture by using a functional integral type of argument,
as explained more fully in \cite{bczsg05}.

The sine-Gordon spectrum contains bound states (breathers), and it
is interesting to calculate their transmission factors. This much
has been done \cite{bczsg05}. One interesting fact is that the
`transmission factor' for the lightest breather has precisely the
same form as the transmission factor in the linearised version of
the model (the exercise set earlier). This strongly suggests it
would also be interesting to attempt to match these breather
transmission factors to perturbative calculations. However, this has
not yet been done.

There are also open questions concerning how to treat defects in
motion. From a classical perspective it seems quite natural that
defects might move and scatter \cite{bczsg05}, however it is less
clear how to describe this in the quantum field theory, although an
attempt has already been made to do so, or indeed to understand what
these objects really are. For example could they also be described
by local fields? Might they correspond to objects that are limiting
cases in the standard theory? For example, it is well-known there is
no two-soliton solution where the two solitons share the same
rapidity, yet in the standard theory their rapidities can be
arbitrarily close.

\section{The affine Toda field theories}

This section will focus on a subset of affine Toda field theories
\cite{Arinshtein79}, namely those associated with the root data of
the Lie algebras $a_r$. Apart from having the most symmetrical
root/weight systems, these are the models for which classically
integrable defects have been described in detail \cite{bcztoda},
whose complex solitons are easy to describe \cite{Hollowood92}, and
whose full set of S-matrices are relatively easy to calculate using
the bootstrap \cite{Hollowood93}.

\label{classicalATFT}

 In the bulk, $-\infty <x<\infty$, an affine Toda field theory
corresponding to the root data of the Lie algebra $a_r$ is
 described conveniently by the Lagrangian density
\begin{equation}\label{affineTodaL}
{\cal L} =\frac{1}{2} \, \partial_{\mu}\phi\cdot
\partial^{\mu}\phi-\frac{m^{2}}
 {{\beta}^{2}}\sum_{j=0}^{r}\, (e^{{\beta}\alpha_{j}\cdot\phi}-1),
\end{equation}
where $m$ and ${\beta}$ are constants, and $r$ is the rank of the
algebra. The vectors $\alpha_j$ with $j=1,\dots,r$ are simple roots
(with the convention $|\alpha_j|^2=2$), and $\alpha_0$ is the lowest
root, defined by
$$\alpha_0=-\sum_{j=1}^{r}\,\alpha_j.$$ The field
$\phi=(\phi_1,\phi_2,\, \dots,\, \phi_r)$ takes values in the
$r$-dimensional Euclidean space spanned by the simple roots
$\{\alpha_j\}$. The extra root $\alpha_0$ distinguishes between  the
massive affine and the massless non-affine Toda field theories. The
massive affine theories are  integrable, possessing infinitely many
conserved charges, a Lax pair representation, and many other
interesting properties, both classically and in the quantum domain.
The simplest choice ($r=1$)  coincides with the sinh-Gordon model.
For further details concerning the affine Toda field theories, see
\cite{Arinshtein79} where further references can be found.

After quantisation, provided the coupling constant ${\beta}$ is
real, and the fields are restricted to be real, the $a_r$ affine
Toda field theory describes $r$ interacting scalars, also known as
fundamental Toda particles, whose classical mass parameters are
given by
\begin{equation}\label{Todaparticlemasses}
m_a=2\,m\sin\left(\frac{\pi a }h\right), \quad a=1,2\dots,r,
\end{equation}
where $h=r+1$ is the Coxeter number of the algebra. On the other
hand, if the fields are permitted to be complex each affine Toda
field theory possesses classical `soliton' solutions
\cite{Hollowood92}. Conventionally, complex affine Toda field
theories are described by the Lagrangian density \eqref{affineTodaL}
in which the coupling constant $\beta$ is replaced with $i \beta$.
Once complex fields are allowed it is clear that the potential
appearing in the Lagrangian density \eqref{affineTodaL} vanishes
whenever the field $\phi$ is constant and equal to
\begin{equation}\label{}
\phi=\frac{2\pi \,w}{\beta}\quad\mbox{with}\quad \alpha_j\cdot w\in
{\bf Z}, \quad\mbox{i.e.}\quad w\in\Lambda_W(a_{r}),
\end{equation}
where $\Lambda_W(a_{r})$ is the weight lattice of the Lie algebra
$a_r$. These constant field configurations have zero energy and
correspond to stationary points of the affine Toda potential.
Soliton solutions smoothly interpolate between these vacuum
configurations as $x$ runs from $-\infty$ to $\infty$. It is natural
to define the `topological charges' characterizing such solutions as
follows:
\begin{equation}\label{topologicalcharges}
Q=\frac{\beta}{2\pi}\int^{\infty}_{-\infty}dx\,\partial_x\phi=
\frac{\beta}{2\pi}\left[\phi(\infty,t) -\phi(-\infty,t)\right],
\end{equation}
and these lie in the weight lattice $\Lambda_W(a_r)$. Assuming
$\phi(-\infty,t)=0$, static solitons may be found for which
$\phi(\infty,t)$ lies in a subset of the weight lattice. In
particular, there are static solutions corresponding to weights
within each of the representations with highest weight $w_a,\
a=1,\dots,r$, satisfying
\begin{equation}
\alpha_i\cdot w_a=\delta_{i a},\quad i,a = 1,\dots, r.
\end{equation}
Explicitly boosted solutions of this type that correspond to the
representation labelled by $a$ have the form
\begin{equation}\label{todasoliton}
\phi^{(a)}=\frac{m^2i}{\beta}\sum^{r}_{j=0}\alpha_{j}\ln \left(1+
E_a\,\omega^{aj}\right), \quad E_a=e^{a_a x-b_a t +\xi_a},\quad
\omega=e^{2\pi i/h},
\end{equation}
where $(a_a, b_a)=m_{a}\, (\cosh{\theta},\sinh{\theta})$, $\xi_a$ is
a complex parameter, and $\theta$ is the soliton rapidity. Despite
the solutions \eqref{todasoliton} being complex, Hollowood
\cite{Hollowood92} showed their total energy and momentum is
actually real and requires masses for  static single solitons
proportional to the mass parameters of the real scalar theory. These
are given by
\begin{equation}\label{singlesolitonmasses}
M_a=\frac{2\,h\,m_a}{\beta^2}, \quad a=1,2\dots,r.
\end{equation}
Moreover, for each $a=1,\dots, r$ there are several solitons whose
topological charges lie in the set of weights of the fundamental
$a^{th}$ representation of $a_r$ \cite{McGhee94}. However, apart
from the two extreme cases, $a=1$ and $a=r$, not every weight
belonging to one of the other representations corresponds to a
static soliton (still something of a mystery). The number of
possible charges for the representation with label $a$ is exactly
equal to the greatest common divisor of $a$ and $h$, the relevant
weights being orbits of the Coxeter element, and explicit
expressions for them may be found in \cite{McGhee94}. The parameter
$\xi_a$ is almost arbitrary but clearly has to be chosen so that
there are no singularities in the solution as $x,t$ vary; shifting
$\xi_a$ by $2\pi ia/h$ changes the topological charge. For the two
extreme representations (with $a=1$ or $a=r$), it is clear repeated
use of this translation steps the charges successively through the
full set of weights.

There are several types of integrable defect for $a_r$ affine Toda
field theory and the distinctions between them are explained in
\cite{bcztoda}. A single defect located at $x=0$ may be described by
the following modified Lagrangian density
\begin{equation}\label{affineTodadefectL}
{\cal L}_d=\theta(-x){\cal L}_\phi +\theta(x){\cal
L}_\psi+\delta(x){\cal D}(\phi,\psi),
\end{equation}
with
\begin{equation}{\cal D}(\phi,\psi)=\frac{1}{2}(\phi\cdot E\partial_t\phi +\phi\cdot
D\partial_t\psi - \partial_t\phi\cdot D\psi
    +\psi\cdot E\partial_t\psi) -{\cal B}(\phi , \psi ),
\end{equation}
where $E$ is an antisymmetric matrix, $D=1-E$,
\begin{equation}
{\cal L}_\phi =\frac{1}{2} \,\partial_{\mu}\phi\cdot
\partial^{\mu}\phi+\frac{m^{2}}
 {\beta^{2}}\,\sum_{j=0}^{r}\, (e^{i\beta\alpha_{j}\cdot\phi}-1),
\end{equation}
and
\begin{equation}\label{affineTodadefectpotential}
   {\cal B}=-\frac{m}{\beta^2}\sum_{j=0}^r\left(\sigma\,
  e^{i\beta\alpha_j\cdot(D^T\phi+D\psi)/2}+
 \frac{1}{\sigma}\, e^{i\beta\alpha_j \cdot D(\phi
 -\psi)/2}\right).
\end{equation}
Here, $\phi$ and $\psi$ are the fields on the left and on the right
of the defect, respectively, and $\sigma$ is the defect parameter.
The matrix $D$ satisfies the following constraints
\begin{equation}\label{constraintsonD}
\alpha_k\cdot D\alpha_j=\left\{%
\begin{array}{ll}
    \phantom{-}2 & \hbox{$k=j$,} \\
    -2 & \hbox{$k=\pi(j)$,} \\
    \phantom{-}0 & \hbox{otherwise,} \\
\end{array}%
\right.\qquad D+D^T=2,
\end{equation}
where $\pi(j)$ indicates a permutation of the simple roots. Choosing
the `clockwise' cyclic permutation,
\begin{equation}\label{permutation}
\nonumber \alpha_{\pi(j)}=\alpha_{j-1},\ j=1,\dots, r,\quad
\alpha_{\pi(0)}=\alpha_r,
\end{equation}
the set of constraints \eqref{constraintsonD} is satisfied by the
 choice,
\begin{equation}\label{Dchoice}
D=2\sum_{a=1}^r w_a\left( w_a-w_{a+1}\right)^T,
\end{equation}
where the vectors $w_a,\ a=1,\dots ,r$ are the fundamental highest
weights of the Lie algebra $a_r$,
 with the added convention
$w_0\equiv w_{r+1}=0$.  Note, the `anticlockwise' cyclic permutation
used in \cite{bcztoda} is effected by substituting the matrix
\eqref{Dchoice} by its transpose.

Given the modified Lagrangian density \eqref{affineTodadefectL} the
corresponding equations of motion and defect conditions are,
respectively,
\begin{eqnarray}\label{equationsofmotion}
\partial^{2}\phi=\frac{m^2i}{\beta}\,
\sum_{j=0}^{r}\,\alpha_j\,e^{i\beta\alpha_{j}\cdot\phi}\quad &x<0, \nonumber \\
\partial^{2}\psi=\frac{m^2i}{\beta}\,
\sum_{j=0}^{r}\,\alpha_j\,e^{i\beta\alpha_{j}\cdot\psi}\quad &x>0,
\end{eqnarray}

\begin{eqnarray}\label{boundaryconditions}
\partial_{x}\phi-E \pt \phi-D \pt \psi+\partial_\phi \cb=0
 \quad &x=0, \nonumber \\
\partial_{x}\psi-D^{T}\pt \phi +E\pt \psi
-\partial_\psi \cb=0 \quad&x=0.
\end{eqnarray}

Many basic properties of \eqref{boundaryconditions} have been noted
elsewhere \cite{bcztoda}\cite{cz07}. Shifting the fields $\phi,\
\psi$ by roots yields another solution with the same energy and
momentum. This is because both the bulk and defect potentials are
invariant under the translations
\begin{equation}
\phi\rightarrow \phi +2\pi r/\beta ,\quad \psi\rightarrow \psi+2\pi
s/\beta,
\end{equation}
where $r,s$ are any two elements of the root lattice. In particular,
constant fields
\begin{equation}\label{staticconfigurations}
(\phi,\psi)=2\pi (r,s)/\beta
\end{equation}
 all have the same
energy and momentum despite having a discontinuity at the location
of the defect. Writing $\sigma=e^{-\eta}$, the energy-momentum of
each of these configurations is
\begin{equation}\label{rootroot}
({\cal E}_0,\, {\cal P}_0)=-\frac{2hm}{\beta^2}(\cosh\eta,\
-\sinh\eta ).
\end{equation}
Other constant configurations are possible and, because of the
invariance under translations by roots, it is enough to consider
configurations $(\phi,\psi)=2\pi (w_p,w_q)/\beta$, where $w_p,w_q$
are fundamental highest weights. As was the case for the sine-Gordon
model there is a conserved momentum associated with the defect.

The system described by the Lagrangian density
\eqref{affineTodadefectL} is neither invariant under parity nor
under time reversal.  By convention, a soliton with positive
rapidity will travel from the left to the right and, at some time,
it will meet the defect located at $x=0$. The soliton $\psi$
emerging on the right will be similar to $\phi$, but  delayed. It is
described by,
\begin{equation}\label{solitonsolution}
\psi^{(a)}=\frac{m^2i}{\beta}\sum^{r}_{j=0}\alpha_{j}\ln \left(1+
z_a\,E_a\,\omega^{aj}\right).
\end{equation}
The expression for the delay $z_a$ was derived in \cite{bcztoda} for
the `anticlockwise' permutation. To obtain the delay for the present
situation it is enough to send the $a^{\rm th}$ soliton to the
$(h-a)^{\rm th}$ soliton in the formula appearing in \cite{bcztoda}.
Therefore the delay is given by
\begin{equation}\label{delay}
z_a=\left(\frac{e^{\,-(\theta-\eta)}+i\,e^{-i\gamma_a}}
{e^{\,-(\theta-\eta)}+i\,e^{\,i\gamma_a}}\right), \quad
\gamma_a=\frac{\pi \,a}{h}.
\end{equation}
The delay is generally  complex with  exceptions being
self-conjugate solitons, corresponding to $a=h/2$ (with $r$ odd),
for which the delay is real. In such cases, the delay is equal to
the delay found for the sine-Gordon model \cite{bczlandau} described
earlier.

Note also that the delays experienced by a soliton, labelled $a$,
and its associated antisoliton, labelled $\bar a = h-a$, are complex
conjugates since $z_{\bar{a}}=\bar{z}_{a}$. For this reason,
solitons and antisolitons are expected to behave differently as they
pass  a defect.

It was also pointed out in \cite{cz07} that the argument of the
phase of the delay \eqref{delay} is given by
\begin{equation}\label{delayarg}
\tan(\mbox{arg}\, z_a)=-\left(\frac{\sin
2\gamma_a}{e^{-2(\theta-\eta)} + \cos2\gamma_a}\right),
\end{equation}
implying that the phase shift produced by the defect can vary
between zero (as $\theta\rightarrow -\infty$) and $-2\gamma_a$ (as
$\theta \rightarrow \infty$), decreasing if necessary through
$-\pi/2$ if $\cos2\gamma_a <0$. On the other hand, the boundaries
between the different topological charge sectors in terms of the
imaginary part of $\xi_a$ (eq\eqref{todasoliton}) are separated by
exactly $2\gamma_a$. This means that a soliton might convert  to one
of the adjacent solitons  as it passes the defect provided
$\mbox{arg}\, z_a$ is sufficiently large. In effect, the defect
imposes a rather severe selection rule on the possible topological
charges of the emerging soliton. In the quantised theory, it is
expected that either the transition matrix has zeroes to reflect
this selection rule, or severely suppressed matrix elements to
represent tunnelling between classically disconnected
configurations. In the sine-Gordon model such an effect was never
evident because the basic representation includes just two states
and transitions between them are always permitted.

The delay \eqref{delay} diverges when
\begin{equation}\label{boundstate}
\theta = \eta + \frac{i\pi}{2}\left(1-\frac{2a}{h}\right),
\end{equation}
and, with the exception of self-conjugate solitons having $a=h/2$
(including the sine-Gordon model where $(a,h)=(1,2)$), this implies
a soliton with real rapidity cannot be absorbed by a defect. For the
sine-Gordon model it was noted already that a classical defect can
absorb a soliton and, within the quantum theory, this phenomenon
implies the existence of unstable bound states. Once the affine Toda
field theories are quantised, however, poles in locations given by
\eqref{boundstate} may correspond to additional states that possess
no classical counterpart. The positions of the poles are expected to
depend on the coupling and it might be the case that there is a
range of couplings for which a bound state exists without the range
including the classical limit.  A phenomenon rather like this does
actually occur in the $a_2$ model and is reported in detail in
\cite{cz07}.

More generally, the delay \eqref{delay} satisfies a classical
bootstrap \cite{cz07} in the sense that when two particles $a,b$ in
the real quantum field theory have a bound state $\bar c$ the
corresponding pole in their S-matrix will occur at rapidities
\begin{equation}
\theta_a=\theta_c-i\bar{U}_{ac}^b,\quad \theta_b= \theta_c
+i\bar{U}_{bc}^a,
\end{equation}
and the corresponding delays \eqref{delay} in the complex classical
theory satisfy
\begin{equation}
z_a(\theta-i\bar{U}_{ac}^b)\, z_b(\theta +i\bar{U}_{bc}^a)=z_{\bar
c}(\theta).
\end{equation}

These observations motivated a study of the triangular equations
\eqref{STT}  in the context of the $a_2$ affine Toda field theory.
This is already substantially more intricate than the similar
analysis for the sine-Gordon model since there are a number of
independent solutions that need to be matched to the Lagrangian
starting point via suitable semi-classical arguments and some that
need to be discarded. The solutions together with their
interpretation, an analysis of the bootstrap and its compatibility
with the triangular relations, an investigation of bound states and
breather transmission factors may all be found in ref\cite{cz07}.
Unfortunately there is no room for these details here.


\section{Concluding remark}

It is quite remarkable that the simple-looking question concerning
integrable shocks asked at the beginning has led to an interesting
avenue of enquiry. The specific question does not appear to have
been explored previously, yet it links with earlier results, such as
(\ref{KL}),  and it is not yet exhausted. The next steps will be to
start a classification of the triangular compatibility relations,
first in the context of other $a_r$ affine Toda field theories, then
afterwards more generally. The transmission matrices are infinite
dimensional, with components labelled by roots, as far as the
defects are concerned, and weights for the solitons, and it will be
interesting to see the variety of possibilities and how they
contrive to match the classical data (if indeed they do). There are
several puzzles to be resolved, one of them being that the $a_r$
models appear to be special in the defect context \cite{bcztoda}. In
all models it remains to be seen if the defects themselves can be
regarded consistently as scattering states. This appears to be
entirely plausible in the sine-Gordon case, but not yet explored for
any other systems.

%


\section*{Acknowledgements}

I am indebted to my colleagues Peter Bowcock and Cristina Zambon for
many useful discussions and several fruitful collaborations.



\begin{thebibliography}{10}

\bibitem{Goodman2002}
R.~H. Goodman, P.~J.~Holmes and M.~I.~Weinstein,
Interaction of sine-Gordon kinks with defects: phase transport
in a two mode model,
{\it Physica D} {\bf 161} (2002), 21.
\newline
R.~H. Goodman, P.~J.~Holmes and M.~I.~Weinstein,
Strong NLS soliton-defect interactions,
{\it Physica D}
{\bf 192} (2004) 215.

\bibitem{Delf94}
G.~Delfino, G.~Mussardo and P.~Simonetti, Statistical models with a
line of defect, {\it Phys.\ Lett.\ } {\bf B328} (1994) 123.
 \newline
G.~Delfino, G.~Mussardo and P.~Simonetti, Scattering theory and
correlation functions in statistical models with a line of defect,
{\it Nucl. Phys. } {\bf B432} (1994) 518.



\bibitem{Konik97}
  R.~Konik and A.~LeClair,
  {Purely transmitting defect field theories},
  {\it Nucl.\ Phys.\ } {\bf B538} (1999) 587.




\bibitem{Mintchev02}
M.~Mintchev, E.~Ragoucy and P.~Sorba, Scattering in the presence of
a reflecting and transmitting impurity, {\it Phys.\ Lett.\ } {\bf
B547} (2002) 313.
\newline
V. Caudrelier, M. Mintchev, E. Ragoucy, Solving the quantum
non-linear Schrodinger equation with delta-type impurity, {\it
J.Math.Phys.} {\bf 46} (2005) 042703.



\bibitem{bczlandau} P. Bowcock, E. Corrigan and C. Zambon,
Classically integrable field theories with defects,  in
 {\it Int.
J. Mod. Physics} {\bf A19} (Supplement) (2004) 82.

\bibitem{bcztoda} P. Bowcock, E. Corrigan and C. Zambon,
Affine Toda field theories with defects, {\it JHEP} {\bf 01} (2004)
056.



\bibitem{cz04}E. Corrigan and C. Zambon, Aspects of sine-Gordon solitons, defects
and gates, {\it J. Phys. A: Math. Gen.}{\bf\ 37} (2004) L471-L477.


\bibitem{bczsg05} P. Bowcock, E. Corrigan and C. Zambon,
Some aspects of jump-defects in the quantum sine-Gordon model,  {\it
JHEP} {\bf 08} (2005) 023.

\bibitem{cz06}
E. Corrigan and C. Zambon, Jump-defects in the nonlinear
Schr\"odinger model and other non relativistic field theories, {\it
Nonlinearity} {\bf 19} (2006) 1447-1469.

\bibitem{cz07}E. Corrigan and  C. Zambon, On purely transmitting defects in affine Toda field theories,
{\it JHEP} {\bf 07} (2007)001.

\bibitem{caudrelier07}
V. Caudrelier, On a systematic approach to defects in classical
integrable field theories, { arXiv:0704.2326} [math-ph]

\bibitem{Gomes06}
J.F. Gomes, L.H. Ymai and A.H. Zimerman: Classical Integrable Super
sinh-Gordon equation with defects, {\it J. Phys. A} \textbf{39}
(2006) 7471.
\newline
J.F. Gomes, L.H. Ymai, A.H. Zimerman, Integrablility of a Classical
$N= 2$ Super Sinh-Gordon Model with Jump Defects, arXiv:0710.1391[hep-th]

\bibitem{Backlund}
C. Rogers and W.K. Schief: \textit{B\"acklund and Darboux
Transformations: Geometry and Modern Applications in Soliton
Theory}, Cambridge Text in Applied Mathematics, Cambridge
University Press 2002.
\bibitem{bcdr}
P. Bowcock, E. Corrigan, P.E. Dorey and R.H. Rietdijk, Classically
integrable boundary conditions for affine Toda field theories, {\it
Nucl. Phys.} \textbf{B445} (1995) 469.

\bibitem{Scott73}
A.C. Scott, F.Y.F. Chu and D.W. McLaughlin,
The soliton: a new concept in applied science,
{\it IEEE Proc.} \textbf{61} (1973) 1443.
\newline
P. J. Drazin and R. S. Johnson, {\em Solitons: an
introduction}, Cambridge University Press 1989.


\bibitem{Zams79} A. Zamolodchikov and Al. Zamolodchikov, Factorized S-Matrices
in Two Dimensions as the Exact solutions of Certain Relativistic
Quantum Field Theory Models, {\it Ann. Phys.} {\bf 120} (1979) 253.

\bibitem{Jimbo89}
M. Jimbo, Introduction to the Yang-Baxter equation, {\it Int. J.
Mod. Phys.} {\bf A4} (1989) 3759.

\bibitem{Arinshtein79} A. E. Arinshtein, V. A. Fateev and A. B.
Zamolodchikov, Quantum S-matrix of the $(1+1)$-dimensional Toda
chain, {\it  Phys. Lett.} {\bf B87} (1979) 389;
\newline A. V. Mikhailov, M. A. Olshanetsky and A. M.
Perelomov, Two-dimensional generalized Toda lattice, {\it Commun.
Math. Phys.} {\bf 79} (1981) 473;
\newline G. Wilson, The modified Lax and two-dimensional Toda
lattice equations associated with simple Lie algebras, {\it  Ergod.
Th. and Dynam. Sys.} {\bf 1} (1981) 361.
\newline
  H. W. Braden, E. Corrigan, P. E. Dorey and R. Sasaki,
  Affine toda field theory and exact S matrices,
  {\it Nucl. Phys. } {\bf B338} (1990) 689.
\newline
E. Corrigan, Recent developments in affine Toda quantum field
theory, published in \emph{Particle and Fields} (Banff, AB, 1994),
CRM Ser. Math. Phys. Springer, New York, (1999).

\bibitem{Hollowood92} T. J. Hollowood, Solitons in affine Toda field
theories, {\it  Nucl. Phys.} {\bf B384} (1992) 523;

\bibitem{Hollowood93} T. J. Hollowood, Quantizing $Sl(N)$ solitons and the
Hecke Algebra, {\it Int. J. Mod. Phys.} {\bf A8} (1993) 947.

\bibitem{McGhee94}
 W. A. McGhee, The Topological Charges of the
$a_n^{(1)}$ Affine Toda Solitons, {\it Int. J. Mod. Phys.} {\bf A9}
(1994) 2645.



\end{thebibliography}
\end{document}